\documentclass[preprint]{aastex61}
\usepackage[hyphenbreaks]{breakurl}
\usepackage[latin1]{inputenc}
\submitjournal{ApJ}

\shorttitle{Galactic cosmic-ray flux depressions observed with LISA Pathfinder}
\shortauthors{Armano et al.}

\begin{document}

\title {Characteristics and energy dependence of recurrent galactic cosmic-ray flux depressions and of a Forbush decrease  with LISA Pathfinder}

\correspondingauthor{Catia Grimani}
\email{catia.grimani@uniurb.it}

\author{M. Armano}
\affiliation{European Space Astronomy Centre, European Space Agency, Villanueva de la Ca\~{n}ada, 28692 Madrid, Spain}

\author{H. Audley}
\affiliation{Albert-Einstein-Institut, Max-Planck-Institut f\"ur Gravitationsphysik und Leibniz Universit\"at Hannover, Callinstra{\ss}e 38, 30167 Hannover, Germany}

\author{J. Baird}
\affiliation{High Energy Physics Group, Physics Department, Imperial College London, Blackett Laboratory, Prince Consort Road, London, SW7 2BW, UK}

\author{M. Bassan}
\affiliation{Dipartimento di Fisica, Universit\`a di Roma ``Tor Vergata'' and INFN sezione Roma Tor Vergata, 00133 Roma, Italy}

\author{S. Benella}
\affil{DISPEA, Universit\`a di Urbino ``Carlo Bo'', Via S. Chiara, 27, 61029 Urbino, Italy}
\affil{INFN - Sezione di Firenze, via G. Sansone 1, 50019, Sesto Fiorentino, Firenze, Italy}

\author{P. Binetruy}
\altaffiliation{Deceased 1 April 2017}
\affiliation{APC, Univ Paris Diderot, CNRS/IN2P3, CEA/lrfu, Obs de Paris, Sorbonne Paris Cit\'e, France}

\author{M. Born}
\affiliation{Albert-Einstein-Institut, Max-Planck-Institut f\"ur Gravitationsphysik und Leibniz Universit\"at Hannover, Callinstra{\ss}e 38, 30167 Hannover, Germany}

\author{D. Bortoluzzi}
\affiliation{Department of Industrial Engineering, University of Trento, via Sommarive 9, 38123 Trento, and Trento Institute for Fundamental Physics and Application / INFN}

\author{A. Cavalleri}
\affiliation{Istituto di Fotonica e Nanotecnologie, CNR-Fondazione Bruno Kessler, I-38123 Povo, Trento, Italy}

\author{A. Cesarini}
\affil{DISPEA, Universit\`a di Urbino ``Carlo Bo'', Via S. Chiara, 27, 61029 Urbino, Italy}

\author{A. M. Cruise}
\affiliation{The School of Physics and Astronomy, University of Birmingham, Birmingham, UK}

\author{K. Danzmann}
\affiliation{Albert-Einstein-Institut, Max-Planck-Institut f\"ur Gravitationsphysik und Leibniz Universit\"at Hannover, Callinstra{\ss}e 38, 30167 Hannover, Germany}

\author{M. de Deus Silva}
\affiliation{European Space Astronomy Centre, European Space Agency, Villanueva de la Ca\~{n}ada, 28692 Madrid, Spain}

\author{I. Diepholz}
\affiliation{Albert-Einstein-Institut, Max-Planck-Institut f\"ur Gravitationsphysik und Leibniz Universit\"at Hannover, Callinstra{\ss}e 38, 30167 Hannover, Germany}

\author{G. Dixon}
\affiliation{The School of Physics and Astronomy, University of Birmingham, Birmingham, UK}

\author{R. Dolesi}
\affiliation{Dipartimento di Fisica, Universit\`a di Trento and Trento Institute for Fundamental Physics and Application / INFN, 38123 Povo, Trento, Italy}

\author{M. Fabi}
\affil{DISPEA, Universit\`a di Urbino ``Carlo Bo'', Via S. Chiara, 27, 61029 Urbino, Italy}

\author{L. Ferraioli}
\affiliation{Institut f\"ur Geophysik, ETH Z\"urich, Sonneggstrasse 5, CH-8092, Z\"urich, Switzerland}

\author{V. Ferroni}
\affiliation{Dipartimento di Fisica, Universit\`a di Trento and Trento Institute for Fundamental Physics and Application / INFN, 38123 Povo, Trento, Italy}

\author{N. Finetti}
\affiliation{INFN - Sezione di Firenze, via G. Sansone 1, 50019, Sesto Fiorentino, Firenze, Italy}
\affiliation{Dipartimento di Scienze Fisiche e Chimiche, Universit\`a degli Studi dell'Aquila, Via Vetoio, Coppito, 67100 L'Aquila, Italy}

\author{E. D. Fitzsimons}
\affiliation{The UK Astronomy Technology Centre, Royal Observatory Edinburgh, Blackford Hill, Edinburgh, EH9 3HJ, UK}

\author{M. Freschi}
\affiliation{European Space Astronomy Centre, European Space Agency, Villanueva de la Ca\~{n}ada, 28692 Madrid, Spain}

\author{L. Gesa}
\affiliation{Institut de Ci\`encies de l'Espai (CSIC-IEEC), Campus UAB, Carrer de Can Magrans s/n, 08193 Cerdanyola del Vall\`es, Spain\\
Institut d'Estudis Espacial de Catalunya (IEEC), Edifici Nexus I, C/ Gran Capit\`a 2-4, despatx 201, E-08034 Barcelona, Spain}

\author{F. Gibert}
\affiliation{Dipartimento di Fisica, Universit\`a di Trento and Trento Institute for Fundamental Physics and Application / INFN, 38123 Povo, Trento, Italy}

\author{D. Giardini}
\affiliation{Institut f\"ur Geophysik, ETH Z\"urich, Sonneggstrasse 5, CH-8092, Z\"urich, Switzerland}

\author{R. Giusteri}
\affiliation{Dipartimento di Fisica, Universit\`a di Trento and Trento Institute for Fundamental Physics and Application / INFN, 38123 Povo, Trento, Italy}

\author[0000-0002-5467-6386]{C. Grimani}
\affiliation{DISPEA, Universit\`a di Urbino ``Carlo Bo'', Via S. Chiara, 27, 61029 Urbino, Italy}
\affiliation{INFN - Sezione di Firenze, via G. Sansone 1, 50019, Sesto Fiorentino, Firenze, Italy}

\author{J. Grzymisch}
\affiliation{European Space Technology Centre, European Space Agency, Keplerlaan 1, 2200 AG Noordwijk, The Netherlands} 
\author{I. Harrison}
\affiliation{European Space Operations Centre, European Space Agency, 64293 Darmstadt, Germany}

\author{G. Heinzel} 
\affiliation{Albert-Einstein-Institut, Max-Planck-Institut f\"ur Gravitationsphysik und Leibniz Universit\"at Hannover, Callinstra{\ss}e 38, 30167 Hannover, Germany}

\author{M. Hewitson}
\affiliation{Albert-Einstein-Institut, Max-Planck-Institut f\"ur Gravitationsphysik und Leibniz Universit\"at Hannover, Callinstra{\ss}e 38, 30167 Hannover, Germany}

\author{D. Hollington}
\affiliation{High Energy Physics Group, Physics Department, Imperial College London, Blackett Laboratory, Prince Consort Road, London, SW7 2BW, UK}

\author{D. Hoyland}
\affiliation{The School of Physics and Astronomy, University of Birmingham, Birmingham, UK}

\author{M. Hueller}
\affiliation{Dipartimento di Fisica, Universit\`a di Trento and Trento Institute for Fundamental Physics and Application / INFN, 38123 Povo, Trento, Italy}

\author{H. Inchausp\'e}
\affiliation{APC, Univ Paris Diderot, CNRS/IN2P3, CEA/lrfu, Obs de Paris, Sorbonne Paris Cit\'e, France}

\author{O. Jennrich}
\affiliation{European Space Technology Centre, European Space Agency, Keplerlaan 1, 2200 AG Noordwijk, The Netherlands}

\author{P. Jetzer}
\affiliation{Physik Institut, Universit\"at Z\"urich, Winterthurerstrasse 190, CH-8057 Z\"urich, Switzerland}

\author{N. Karnesis}
\affiliation{Albert-Einstein-Institut, Max-Planck-Institut f\"ur Gravitationsphysik und Leibniz Universit\"at Hannover, Callinstra{\ss}e 38, 30167 Hannover, Germany}

\author{B. Kaune}
\affiliation{Albert-Einstein-Institut, Max-Planck-Institut f\"ur Gravitationsphysik und Leibniz Universit\"at Hannover, Callinstra{\ss}e 38, 30167 Hannover, Germany}

\author{N. Korsakova}
\affiliation{SUPA, Institute for Gravitational Research, School of Physics and Astronomy, University of Glasgow, Glasgow, G12 8QQ, UK}

\author{C. J. Killow}
\affiliation{SUPA, Institute for Gravitational Research, School of Physics and Astronomy, University of Glasgow, Glasgow, G12 8QQ, UK}

\author{M. Laurenza}
\affiliation{INFN - Sezione di Firenze, via G. Sansone 1, 50019, Sesto Fiorentino, Firenze, Italy}
\affiliation{Istituto di Astrofisica e Planetologia Spaziali, INAF, Roma, Italy}

\author{J. A. Lobo}
\altaffiliation{Deceased 30 September 2012}
\affiliation{Institut de Ci\`encies de l'Espai (CSIC-IEEC), Campus UAB, Carrer de Can Magrans s/n, 08193 Cerdanyola del Vall\`es, Spain\\
Institut d'Estudis Espacial de Catalunya (IEEC), Edifici Nexus I, C/ Gran Capit\`a 2-4, despatx 201, E-08034 Barcelona, Spain}

\author{I. Lloro}
\affiliation{Institut de Ci\`encies de l'Espai (CSIC-IEEC), Campus UAB, Carrer de Can Magrans s/n, 08193 Cerdanyola del Vall\`es, Spain\\
Institut d'Estudis Espacial de Catalunya (IEEC), Edifici Nexus I, C/ Gran Capit\`a 2-4, despatx 201, E-08034 Barcelona, Spain}

\author{L. Liu}
\affiliation{Dipartimento di Fisica, Universit\`a di Trento and Trento Institute for Fundamental Physics and Application / INFN, 38123 Povo, Trento, Italy}

\author{J. P. L\'opez-Zaragoza}
\affiliation{Institut de Ci\`encies de l'Espai (CSIC-IEEC), Campus UAB, Carrer de Can Magrans s/n, 08193 Cerdanyola del Vall\`es, Spain\\
Institut d'Estudis Espacial de Catalunya (IEEC), Edifici Nexus I, C/ Gran Capit\`a 2-4, despatx 201, E-08034 Barcelona, Spain}

\author{R. Maarschalkerweerd}
\affiliation{European Space Operations Centre, European Space Agency, 64293 Darmstadt, Germany}

\author{D. Mance}
\affiliation{Institut f\"ur Geophysik, ETH Z\"urich, Sonneggstrasse 5, CH-8092, Z\"urich, Switzerland}

\author{V. Mart\'in}
\affiliation{Institut de Ci\`encies de l'Espai (CSIC-IEEC), Campus UAB, Carrer de Can Magrans s/n, 08193 Cerdanyola del Vall\`es, Spain\\
Institut d'Estudis Espacial de Catalunya (IEEC), Edifici Nexus I, C/ Gran Capit\`a 2-4, despatx 201, E-08034 Barcelona, Spain}

\author{L. Martin-Polo}
\affiliation{European Space Astronomy Centre, European Space Agency, Villanueva de la Ca\~{n}ada, 28692 Madrid, Spain}

\author{J. Martino}
\affiliation{APC, Univ Paris Diderot, CNRS/IN2P3, CEA/lrfu, Obs de Paris, Sorbonne Paris Cit\'e, France}

\author{F. Martin-Porqueras}
\affiliation{European Space Astronomy Centre, European Space Agency, Villanueva de la Ca\~{n}ada, 28692 Madrid, Spain}

\author{I. Mateos}
\affiliation{Institut de Ci\`encies de l'Espai (CSIC-IEEC), Campus UAB, Carrer de Can Magrans s/n, 08193 Cerdanyola del Vall\`es, Spain\\
Institut d'Estudis Espacial de Catalunya (IEEC), Edifici Nexus I, C/ Gran Capit\`a 2-4, despatx 201, E-08034 Barcelona, Spain}

\author{P. W. McNamara}
\affiliation{European Space Technology Centre, European Space Agency, Keplerlaan 1, 2200 AG Noordwijk, The Netherlands} 
\author{J. Mendes}
\affiliation{European Space Operations Centre, European Space Agency, 64293 Darmstadt, Germany}

\author{L. Mendes}
\affiliation{European Space Astronomy Centre, European Space Agency, Villanueva de la Ca\~{n}ada, 28692 Madrid, Spain}

\author{M. Nofrarias}
\affiliation{Institut de Ci\`encies de l'Espai (CSIC-IEEC), Campus UAB, Carrer de Can Magrans s/n, 08193 Cerdanyola del Vall\`es, Spain\\
Institut d'Estudis Espacial de Catalunya (IEEC), Edifici Nexus I, C/ Gran Capit\`a 2-4, despatx 201, E-08034 Barcelona, Spain}

\author{S. Paczkowski}
\affiliation{Albert-Einstein-Institut, Max-Planck-Institut f\"ur Gravitationsphysik und Leibniz Universit\"at Hannover, Callinstra{\ss}e 38, 30167 Hannover, Germany}

\author{M. Perreur-Lloyd}                                                       
\affiliation{SUPA, Institute for Gravitational Research, School of Physics and Astronomy, University of Glasgow, Glasgow, G12 8QQ, UK}

\author{A. Petiteau}
\affiliation{APC, Univ Paris Diderot, CNRS/IN2P3, CEA/lrfu, Obs de Paris, Sorbonne Paris Cit\'e, France}

\author{P. Pivato}
\affiliation{Dipartimento di Fisica, Universit\`a di Trento and Trento Institute for Fundamental Physics and Application / INFN, 38123 Povo, Trento, Italy}

\author{E. Plagnol}
\affiliation{APC, Univ Paris Diderot, CNRS/IN2P3, CEA/lrfu, Obs de Paris, Sorbonne Paris Cit\'e, France}

\author{J. Ramos-Castro}
\affiliation{Department d'Enginyeria Electr\`onica, Universitat Polit\`ecnica de Catalunya,  08034 Barcelona, Spain}

\author{J. Reiche}
\affiliation{Albert-Einstein-Institut, Max-Planck-Institut f\"ur Gravitationsphysik und Leibniz Universit\"at Hannover, Callinstra{\ss}e 38, 30167 Hannover, Germany} 

\author{D. I. Robertson}
\affiliation{SUPA, Institute for Gravitational Research, School of Physics and Astronomy, University of Glasgow, Glasgow, G12 8QQ, UK}

\author{F. Rivas}
\affiliation{Institut de Ci\`encies de l'Espai (CSIC-IEEC), Campus UAB, Carrer de Can Magrans s/n, 08193 Cerdanyola del Vall\`es, Spain\\
Institut d'Estudis Espacial de Catalunya (IEEC), Edifici Nexus I, C/ Gran Capit\`a 2-4, despatx 201, E-08034 Barcelona, Spain}

\author{G. Russano}
\affiliation{Dipartimento di Fisica, Universit\`a di Trento and Trento Institute for Fundamental Physics and Application / INFN, 38123 Povo, Trento, Italy}

\author{F. Sabbatini}
\affiliation{DISPEA, Universit\`a di Urbino ``Carlo Bo'', Via S. Chiara, 27, 61029 Urbino, Italy}

\author{J. Slutsky}
\affiliation{Gravitational Astrophysics Lab, NASA Goddard Space Flight Center, 8800 Greenbelt Road, Greenbelt, MD 20771 USA}

\author{C. F. Sopuerta}
\affiliation{Institut de Ci\`encies de l'Espai (CSIC-IEEC), Campus UAB, Carrer de Can Magrans s/n, 08193 Cerdanyola del Vall\`es, Spain\\
Institut d'Estudis Espacial de Catalunya (IEEC), Edifici Nexus I, C/ Gran Capit\`a 2-4, despatx 201, E-08034 Barcelona, Spain}

\author{T. Sumner}
\affiliation{High Energy Physics Group, Physics Department, Imperial College London, Blackett Laboratory, Prince Consort Road, London, SW7 2BW, UK}

\author{D. Telloni}
\affiliation{INFN - Sezione di Firenze, via G. Sansone 1, 50019, Sesto Fiorentino, Firenze, Italy}
\affiliation{Osservatorio Astrofisico di Torino, INAF, Pino Torinese, Italy}

\author{D. Texier}
\affiliation{European Space Astronomy Centre, European Space Agency, Villanueva de la Ca\~{n}ada, 28692 Madrid, Spain}

\author{J. I. Thorpe}
\affiliation{Gravitational Astrophysics Lab, NASA Goddard Space Flight Center, 8800 Greenbelt Road, Greenbelt, MD 20771 USA}

\author{D. Vetrugno}
\affiliation{Dipartimento di Fisica, Universit\`a di Trento and Trento Institute for Fundamental Physics and Application / INFN, 38123 Povo, Trento, Italy}

\author{S. Vitale}
\affiliation{Dipartimento di Fisica, Universit\`a di Trento and Trento Institute for Fundamental Physics and Application / INFN, 38123 Povo, Trento, Italy}

\author{G. Wanner}
\affiliation{Albert-Einstein-Institut, Max-Planck-Institut f\"ur Gravitationsphysik und Leibniz Universit\"at Hannover, Callinstra{\ss}e 38, 30167 Hannover, Germany}

\author{H. Ward}
\affiliation{SUPA, Institute for Gravitational Research, School of Physics and Astronomy, University of Glasgow, Glasgow, G12 8QQ, UK}

\author{P. Wass}
\affiliation{High Energy Physics Group, Physics Department, Imperial College London, Blackett Laboratory, Prince Consort Road, London, SW7 2BW, UK}

\author{W. J. Weber}
\affiliation{Dipartimento di Fisica, Universit\`a di Trento and Trento Institute for Fundamental Physics and Application / INFN, 38123 Povo, Trento, Italy}

\author{L. Wissel}
\affiliation{Albert-Einstein-Institut, Max-Planck-Institut f\"ur Gravitationsphysik und Leibniz Universit\"at Hannover, Callinstra{\ss}e 38, 30167 Hannover, Germany}

\author{A. Wittchen}
\affiliation{Albert-Einstein-Institut, Max-Planck-Institut f\"ur Gravitationsphysik und Leibniz Universit\"at Hannover, Callinstra{\ss}e 38, 30167 Hannover, Germany}

\author{A. Zambotti}
\affiliation{Department of Industrial Engineering, University of Trento, via Sommarive 9, 38123 Trento, and Trento Institute for Fundamental Physics and Application / INFN}

\author{C. Zanoni}
\affiliation{Department of Industrial Engineering, University of Trento, via Sommarive 9, 38123 Trento, and Trento Institute for Fundamental Physics and Application / INFN}

\author{P. Zweifel}
\affiliation{Institut f\"ur Geophysik, ETH Z\"urich, Sonneggstrasse 5, CH-8092, Z\"urich, Switzerland}



\begin{abstract}

Galactic cosmic-ray (GCR) energy spectra observed in the inner heliosphere are  modulated  by the solar activity, the solar polarity and  structures of solar and  interplanetary  origin. 
A high counting rate particle detector (PD) aboard LISA Pathfinder (LPF), meant for subsystems diagnostics,   was devoted to the measurement of galactic cosmic-ray  and
solar energetic particle  integral fluxes above 70 MeV n$^{-1}$ up to 6500 counts s$^{-1}$. 
 PD data were gathered with a sampling time of 15 s. 
Characteristics and energy-dependence of GCR flux recurrent depressions and of a Forbush decrease dated  August 2, 2016 are reported here.
The capability of interplanetary missions, carrying  PDs for instrument performance purposes, in monitoring the passage of interplanetary coronal mass ejections (ICMEs) is also discussed.

\end{abstract}




\section{Introduction} \label{sec:intro}

 Galactic cosmic-ray (GCR) flux observations in the heliosphere present long-term  ($>$ 1 year) and short-term ($\le$ 27 days) modulations. Both were extensively studied in the last 60 years on Earth with neutron monitors and in space \citep{forb54,forb58,iucci,beer,integral,evenson,ferreira,gri04a,gri07,grim07,bess,kudela,uso11,lau12,lau14,uso17}. 

Long-term variations are associated with the 11-year solar cycle and the 22-year  
 solar polarity  reversal. At solar maximum GCR energy spectra appear depressed  by approximately one
order of magnitude at  100 MeV n$^{-1}$ with respect to similar observations gathered at solar minimum \citep[see for instance][and references therein]{pgs}. 
 Moreover, at solar minimum and during negative (positive) solar polarity periods, defined by the solar magnetic field directed inward (outward) at the Sun north pole, positively (negatively) charged particle fluxes are depressed by a maximum of 40\% at 100 MeV n$^{-1}$ with respect to measurements performed during opposite  periodicities \citep{imp8,gil1}. 
Positively charged particles propagate mainly sunward in the ecliptic along the heliospheric current sheet (HCS) during negative solar polarity periods and over the poles during positive polarity epochs. The opposite holds for negatively charged particles \citep{potlan,ferr}. Particles propagating along the HCS lose more energy than those coming from the poles \citep{strauss}. 

The most intense short-term GCR flux drops occur during classical, non recurrent, Forbush decreases \citep{forb37,cane1}. These depressions are characterized by maximum GCR flux decreases of 30\% at 100 MeV n$^{-1}$ and are associated with the passage of interplanetary counterparts of coronal mass ejections. Recurrent depressions are caused by corotating high-speed solar wind streams \citep[see][for instance]{iucci1}. 
Quasi-periodicities of 27 days, 13.5 days and 9 days, correlated with the Sun rotation period (27.28 days for an Earth observer) and higher harmonics, are  
observed 
in the cosmic-ray flux, in the  solar wind plasma, in the interplanetary  magnetic field and in the geomagnetic activity indices \citep{jasa,emery}. 
These investigations are typically carried out with neutron monitors  that allow for long-term studies of the role of interplanetary structures in modulating the GCR flux  \citep[see for instance][]{sim54,gil2,kudela,budu1}.
A correlation of the GCR flux short-term variations with the BV product of the interplanetary 
magnetic field (IMF) intensity (B) and the solar wind speed (V) was investigated by \citet{sabbah1}. 
This approach  takes into account both cosmic-ray diffusion from interplanetary magnetic field and 
convection in the solar wind. From the point of view of geomagnetic indices, a good correlation 
of A$_p$ and K$_p$ with both  BV and BV$^2$, was found by \citet{sabbah2}.
Depressions of the cosmic-ray flux were  studied  in space since the sixties \citep[see for instance][]{mrb}.  \citet{richetal} carried out an extensive campaign of observations of GCR flux short-term variations above a few tens of MeV 
aboard the  Helios 1, Helios 2 and IMP-8 spacecraft. These observations 
indicated that the effects  of corotating
interaction regions  (CIRs), generated when  high-speed solar wind streams, associated with stable, low-latitude extensions of polar coronal holes, 
 overtake leading slow solar wind from the equatorial regions of the Sun, 
are at the origin of short-term GCR flux modulations \citep[see also][]{richar}.  

A high counting rate particle detector (PD; \citet{parde}) hosted aboard the European Space Agency (ESA)
LISA Pathfinder (LPF) mission \citep{lisapf1,lisapf2,armano}, allowed for the monitoring of the integral proton and helium nucleus
fluxes above 70 MeV n$^{-1}$  \citep{ara,mateos}  with  statistical uncertainty  at percent level on 1-hour binned  data between February 2016 and July 2017.

The energy-dependence of GCR short-term depressions can be studied by exploiting the  contemporaneous measurements of cosmic rays in space 
above a few tens of MeV with missions carrying PDs and on Earth with neutron monitors 
located at different geographic latitudes.
GCR counting rates observed with neutron monitors vary  proportionally to the  
cosmic-ray flux, thus providing a direct measurement of the same, 
at energies larger than the {\it effective energy} \citep{gil3} which ranges between  11-12 GeV and  above 20 GeV for near-polar and equatorial stations, respectively.

This paper reports on the characteristics of GCR flux periodicities and depressions observed 
during the Bartels rotations (BRs) 2490-2508 (from February 18, 2016 through July 3, 2017) after properly taking into account the effects of long-term variations. 
It is recalled here that the BR number corresponds to the number of 27-day rotations of the Sun since
 February 8, 1832.
The years 2016-2017 were characterized by the presence of  near-equatorial coronal holes and equatorward  extensions  of polar coronal holes,
resulting in a very   favourable period to carry out the study illustrated here.
The energy-dependence of recurrent and non-recurrent GCR depressions is also investigated. In  particular, it is reported  on the characteristics of a classical Forbush decrease,  a sudden  depression of the GCR flux observed with LPF  on August 2nd, 2016. This occurrence was  associated with an increase of the IMF intensity due to the passage of an interplanetary coronal mass ejection (ICME; \url{http://www.srl.caltech.edu/ACE/ASC/DATA/level3/icmetable2.htm} and \citet[][]{ricane}) that caused  a  geomagnetic disturbance of modest intensity started at 21.30 UT of the same day.  GCR proton energy spectra in August 2016 before and at the deep of the depression are estimated and presented in this work.
These observations indicate the value of interplanetary missions carrying particle detectors  that, while primarily devoted to mission performance purposes, can also provide valuable measurements for space science and space weather studies \citep[see also][and references therein]{integral,rosetta}. 

 This manuscript is organized as it follows: Section 2 describes the characteristics of the LPF mission. In Section 3 are presented the parameterizations of the proton and helium energy spectra   during the LPF mission.  In Section 4 and Section 5  are reported the characteristics and the energy dependence of the observed GCR flux short-term variations, respectively.  Finally, in Section 6 it is discussed the capability of the LPF PD to monitor the passage of interplanetary coronal mass ejections.

\section{LPF mission and orbit} \label{sec:style}

 LPF was the  technology demonstrator mission for LISA, the first space interferometer devoted  to gravitational wave detection in the
frequency range 10$^{-4}$ Hz - 10$^{-1}$ Hz \citep{lisa}.
The LPF spacecraft  was launched from the Kourou base in  French Guiana on December 3rd, 2015 aboard a Vega rocket.
It reached its final orbit (which took approximately 6 months to complete) around the Earth-Sun Lagrangian point L1 at 1.5 million km from Earth at the end of January  2016.
The LPF orbit was inclined at about 45 degrees with respect to the ecliptic plane.
Orbit minor and major axes were approximately of 0.5 million km and 0.8 million km, respectively. The satellite spinned on its own axis in six months. 
The LPF satellite
carried two, 2-kg cubic platinum-gold free-floating test masses that play the role of mirrors 
of the interferometer. 
Protons and ions of galactic or solar origin with energies larger than 100 MeV n$^{-1}$  penetrated or interacted in about  13 g cm$^{-2}$ of spacecraft and instrument materials charging the LPF test masses.
This charging process    
results in  spurious noise  forces on both test masses \citep{shaul,wass}. 
A PD \citep{parde} was placed aboard LPF for  {\it in situ} monitoring of GCR and solar particle overall flux.
The LPF PD was mounted  behind the spacecraft solar panels with its viewing axis along the Sun-Earth direction.
It consisted of two $\sim$ 300 $\mu$m thick silicon wafers 
of  1.40 x 1.05 cm$^2$ area,   placed in
a telescopic arrangement at a distance of 2 cm. For particle  energies $>$ 100 MeV n$^{-1}$ the instrument  geometrical factor  was  found to be energy independent and
equal to  9 cm$^2$ sr for particle isotropic incidence on each silicon layer.
When particles traversed both silicon wafers within 525 ns of each other (coincidence mode), the geometrical factor was  about one tenth
of this  value. A shielding copper  box of 6.4 mm thickness surrounded the silicon wafers.
 The shielding material stopped particles with energies smaller than 70 MeV 
n$^{-1}$. The PD  allowed for the counting of
 protons and helium nuclei
 traversing  each silicon layer (single counts) and for the measurement of ionization energy losses of particles in coincidence mode.
The single counts were gathered with a sampling time of 15 s and ionization energy losses of events in coincidence mode  were stored  in the form of histograms over periods of 600 seconds and then sent to the on-board computer.
The maximum allowed detector  counting rate
was 6500 counts s$^{-1}$ on both silicon wafers, corresponding to an event integrated  proton fluence of   10$^8$ protons cm$^{-2}$ at
energies $>$ 100 MeV. 
In coincidence mode 5000 energy deposits per second   was the  saturation limit.
The occurrence of SEP events with fluences larger than the saturation limit was estimated  to be less than one per year for the period the LPF spacecraft remained into orbit around L1 \citep{nymmik1,nymmik2,gri12}. 
As a matter of fact, no SEP events  characterized by a proton differential flux
above a few tens of MeV n$^{-1}$  overcoming that of galactic origin were observed during the period of the LPF mission operations considered for this analysis.                      

\section{Galactic cosmic-ray proton and helium nucleus energy spectra during the LPF mission} \label{subsec:tables}

  The LPF  15-s  proton (p) and helium (He) single counts gathered between mid-February 2016 and July 3, 2017 were hourly-averaged in order to limit the statistical uncertainty on the measurements to 1\% (see Fig. 1).  Observations were interrupted only for brief, planned system resets. The GCR count rate appears modulated on time scales of several days and presents an increasing trend over the mission lifetime due to a decreasing level of the solar activity.  
\begin{figure*}
\begin{center}
\includegraphics[width=0.4\textwidth,angle=90]{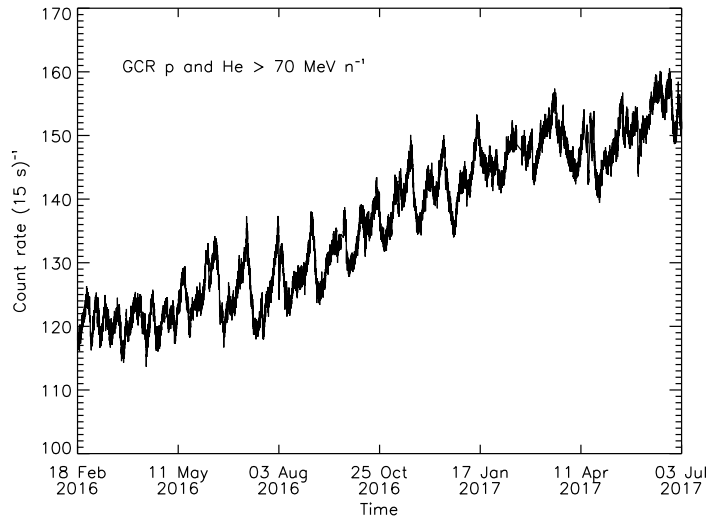}
\caption{Fifteen second hourly-averaged GCR single count rate observed with the PD aboard the LPF mission.}
\label{figure1}
\end{center}
\end{figure*}
It is worthwhile to recall that LPF was sent into orbit during the descending phase of the solar cycle N$^{\circ}$ 24 under a positive polarity period.
In \citet{grim07} it was shown that during positive polarity periods the energy spectra, $J(r,E,t)$, of cosmic 
rays at a distance $r$ from the Sun and at a time $t$,  are well represented by  the
 symmetric model in the $force$ $field$ $approximation$
 by  Gleeson and Axford (G\&A; \citet{gle68}) assuming  time-independent interstellar intensities $J(\infty,E+\Phi)$
and  an energy loss parameter $\Phi$:


\begin{equation}
\frac{J(r,E,t)}{E^2-E^2_0}=\frac{J(\infty,E+\Phi)}{(E+\Phi)^2-E^2_0}
\label{equation1}
\end{equation}


where $E$ and $E_0$ represent  the particle total energy and rest mass, respectively.
For Z=1 particles with rigidity (particle momentum  per unit charge) larger than  100 MV, the effect of the solar activity is completely defined by the $solar$ $modulation$ $parameter$ $\phi$ that, at these energies, is equal to $\Phi$  \citep[see also][]{gri09}.

The solar modulation parameter for the first year of the LPF mission (December 2015 - December 2016) was taken from  \url{http://cosmicrays.oulu.fi/phi/Phi\_mon.txt} \citep[see also][]{uso06}. For the same period, the  GCR single counts per sampling time of 15 seconds, averaged over each  BR (GCR$_{15s}$), were calculated.
A linear correlation was found between the solar modulation parameter $\phi$ and GCR$_{15s}$:

\begin{equation}
 GCR_{15s}=-0.23272\ \phi(MV)+230.73
\label{equation1}
\end{equation}

\noindent as it is shown in Fig. 2 at the right side of the dashed line, indicated by {\it DATA}. 
\begin{figure}
  \begin{center}
  \includegraphics[width=3.5in]{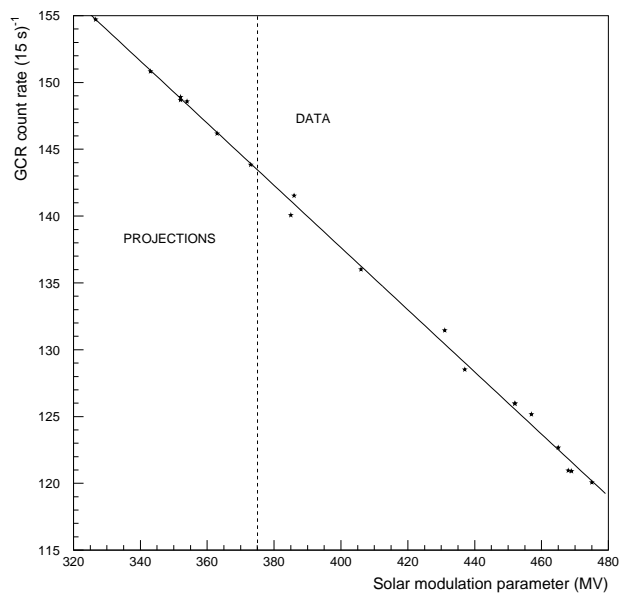}
  \caption{Solar modulation parameter and  LPF PD GCR single count rate in 15 s sampling time averaged over each BR during the LPF mission. High (low) values of the solar modulation parameter correspond to mission beginning (end). See text for details. Estimates of the solar modulation parameter up to December 2016 appear in \url{http://cosmicrays.oulu.fi/phi/Phi\_mon.txt}. In 2017  projections of the solar modulation parameter were carried out on the basis of the parameterization appearing as a continuous line and reported in equation (2).}
  \label{figure1}
\end{center}
  \end{figure}
This observation suggests that  the  LPF PD did not present any detectable loss of efficiency during the first year of the mission lifetime. Therefore, the same was reasonably assumed for the last six months of mission operations. Projections of the solar modulation parameter for the year 2017 (for which estimates are not available in  \url{http://cosmicrays.oulu.fi/phi/Phi\_mon.txt}) were  carried out by extrapolating the same trend shown by GCR$_{15s}$ and  $\phi$  in 2016 (see Fig. 2 at the left of the dashed line indicated by {\it PROJECTIONS}).  The observed PD single count rate increased by more than 20\% during the LPF mission  due to a decreasing  solar activity. The monthly sunspot number (\url{http://www.sidc.be/silso/home}) was observed to decrease smoothly from 58 to 18.5 during the first year of the LPF mission while from January 2017 through the beginning of  July 2017 the sunspot number did not change appreciably varying from 26.1 to 19.4. Therefore,  the value of $\phi$ assumed here at the end of the LPF mission can be considered a lower limit. 

\begin{figure}
  \begin{center}
  \includegraphics[width=3.5in]{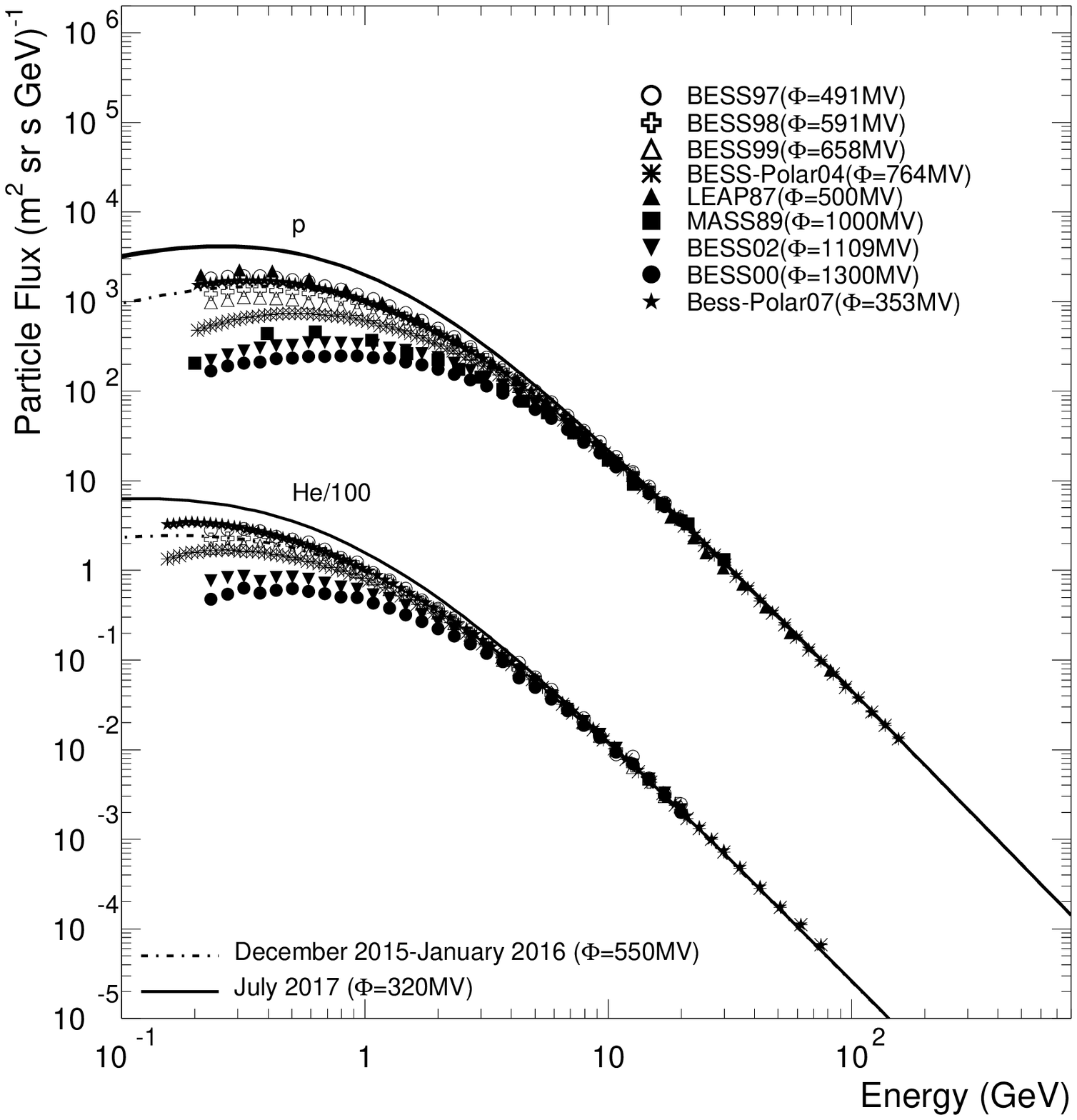}
  \caption{GCR proton and helium energy spectra measurements \citep[][and references therein]{bess}. Estimated energy spectra at the beginning  (December 2015 - January 2016)  at the end of the LPF mission (July 2017) are also indicated as dot-dashed and continuous lines, respectively. The helium flux appears properly scaled in order not to superpose lines.
}
  \label{figure1}
 \end{center}
\end{figure}



The proton  and helium   energy differential fluxes at the beginning (December 2015 - January 2016; $\phi$=550 MV)
and at the    end (July 2017; $\phi$=320 MV) of the LPF mission were estimated with the  model by G\&A by using the proton  and helium  energy spectra
at the interstellar medium obtained from a series of balloon flights of the BESS and BESS-Polar experiments
 \citep[see][for details]{bess,besspolar}. The BESS, BESS-Polar and other balloon-borne experiment data gathered during different periods of solar activity
 and solar polarity, are reported in Fig. 3. In this figure open (solid) symbols indicate data gathered during positive (negative) polarity periods.
In \citet[][]{gri04b} and references therein it was shown that contemporaneous observations of GCR fluxes in the inner heliosphere   show variations of $\sim$ 3\% AU$^{-1}$ and 0.33\% per latitude degree off the ecliptic. It can be concluded that particle spectra at the interstellar medium  obtained with data gathered near Earth can  also be used for LPF which orbited at just 0.01 AU from Earth.  
Finally, the interstellar spectra  by  BESS-BESS-Polar were privileged in this work since they were inferred  from proton and helium  observations gathered during conditions of solar activity and solar polarity similar to those of LPF.


 The energy spectra, F(E), obtained with the G\&A model for LPF were interpolated with the function appearing  in equation 3, which is  well  representative of the GCR observations trend in the inner heliosphere  between a few tens of MeV and hundreds of GeV within experimental errors \citep[see for details][]{pgs}:

\begin{equation}
F(E)= A\ (E+b)^{-\alpha}\ E^{\beta}  \ \ \ {\rm particles\ (m^2\ sr\ s\ GeV\ n^{-1})^{-1}},
\label{equation2}
\end{equation}

\noindent where  $E$ is the particle kinetic energy per nucleon.  The
parameters $A$, $b$, $\alpha$ and $\beta$  were estimated for proton and helium nucleus energy spectra  at the beginning and at the end of the LPF mission and reported in Table 1.
\begin{deluxetable}{ccccc}
\tablecaption{Parameterizations of proton and helium energy spectra at the beginning and at the end of the LPF mission  according to the  function
$F(E)= A\ (E+b)^{-\alpha}\ E^{\beta}  \ \ \ {\rm particles\ (m^2\ sr\ s\ GeV\ n^{-1})^{-1}}$
\label{tabparam}}
\tablehead{
\colhead{} & \colhead{A} & \colhead{b} & \colhead{$\alpha$}& \colhead{$\beta$}
}
\startdata
p (Dec.\ 2015 - Jan.\ 2016)  & 18000 & 1.19 & 3.66 & 0.87\\
p (July\ 2017)  & 18000 & 0.82 & 3.66 & 0.87\\
He (Dec.\ 2015 - Jan.\ 2016) & 850  & 0.96  & 3.23 & 0.48\\
He (July\ 2017) & 850  & 0.68  & 3.23 & 0.48\\
\enddata
\end{deluxetable}
The  energy spectra obtained here for the beginning of the LPF mission (dot-dashed curve in Fig. 3;  $\phi$=550 MV) lie, as expected,  between the BESS97 ($\phi$=491 MV) and the BESS98  ($\phi$=591 MV) data, gathered during a positive polarity period. In the same figure maximum projections of proton and helium energy spectra at the end of the LPF mission appear as continuous lines. 
The same value of  $\phi$ was used for both proton and helium  energy spectra estimates.

\section{Observations  of GCR  flux short-term variations  aboard LPF}

The power spectral density (PSD) from the Lomb-Scargle (LS; \citet{lomb1976,scargle1982}) periodogram analysis of the whole LPF GCR data sample
adopted for this study is shown in Fig. 4.
\begin{figure*}
\begin{center}
\includegraphics[width=0.6\textwidth]{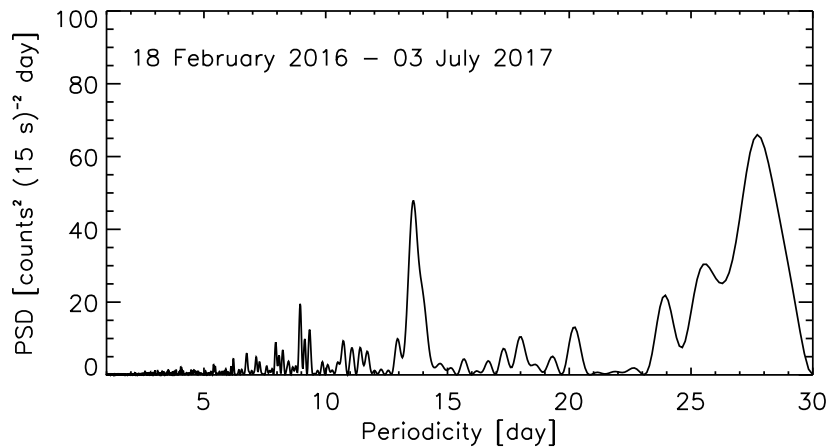}
\caption{Power spectral density  from the Lomb-Scargle periodogram analysis
applied to the whole LPF PD data set adopted in this work. 
Sun rotation and  higher harmonics periodicities are dominant.
}
\label{figure1}
\end{center}
\end{figure*}
The LS  periodogram  technique is used here to retrieve the  periodicities of galactic cosmic-ray modulation. 
Fig. 4 shows that periodicities of 9 days, 13.5 days and 27 days correlated to the Sun rotation and higher harmonics periodicities are present in the whole GCR PD data.
In order to assess the time variability of these dominant periodicities, 
the  period of observations was divided in three sub-intervals, each encompassing about four months and a half.
\begin{figure*}
\begin{center}
\includegraphics[width=0.5\textwidth]{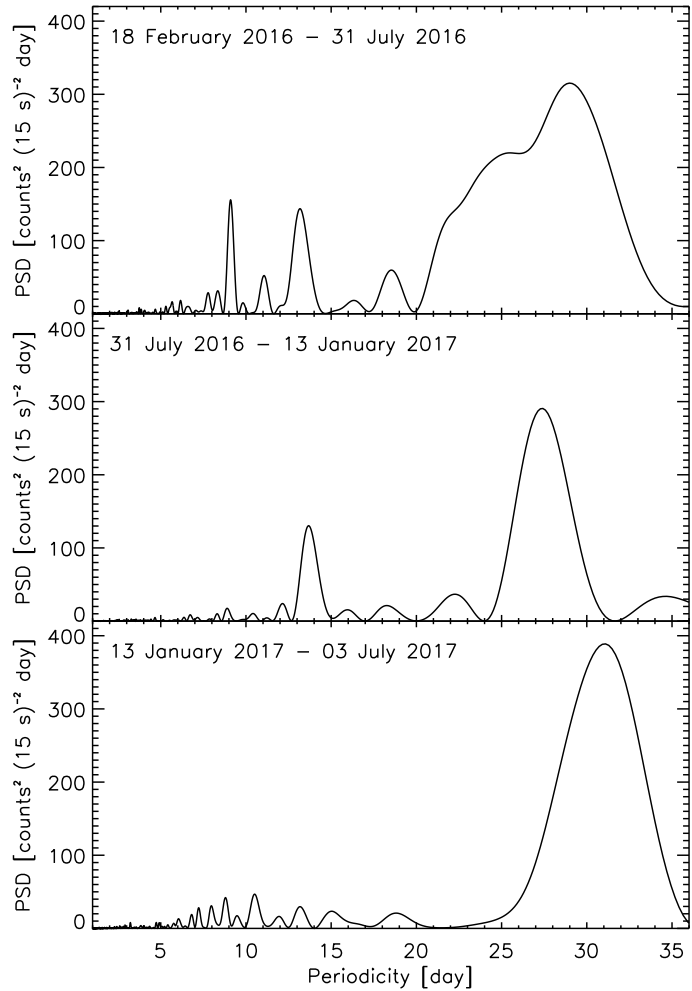}
\caption{Power spectral density from the Lomb-Scargle periodogram analysis applied to the PD data gathered in the time interval indicated in each
 panel (which refers to one third of the whole LPF PD data set). 
Sun rotation and  higher harmonics periodicities are modulated throughout the observational period.
}
\label{figure1}
\end{center}
\end{figure*}

The corresponding LS power spectral densities are displayed in Fig. 5.
The 9-day and 13.5-day periodicities are strongly modulated in time and progressively damped, with the former
 being the first to disappear. The periodicity related to the Sun rotation is present during the whole 
observational period, though its value slightly changes in time from about 27 days (middle panel) to 31 days (third panel).

In order to study the occurrence and the characteristics of individual GCR flux short-term depressions, 
the LPF PD observations during each BR were compared to interplanetary magnetic field, 
solar wind plasma parameters  and to neutron monitor measurements. Moreover, data gathered during each Bartels rotation were compared by eye to those observed during previous and subsequent Bartels rotations  in order to detect the presence of recurring and non-recurring patterns in the variation of GCR data and solar wind parameters.

The effects of the decrease of the solar activity  over the mission
were reduced by considering
the fractional variations of the cosmic-ray flux  with respect
 to the average value during each  BR. 
The same approach was considered in \citet{wied} for the ACE experiment in L1.   Forty-four  recurrent depressions and one classical Forbush decrease were observed.
The commencement of individual depressions was set at the beginning of each continuous decrease of the GCR flux  observed for more than 12 hours.   
GCR flux depressions with duration $>$ 1 day and amplitude $>$ 1.5\%  were considered for this analysis. Small increases and depressions ($<$ 1.5\% in amplitude) lasting less than one day were  at the limit of statistical significance and therefore were disregarded by interpolating the data trend.
\begin{figure}
  \begin{center}
  \includegraphics[width=0.7\textwidth]{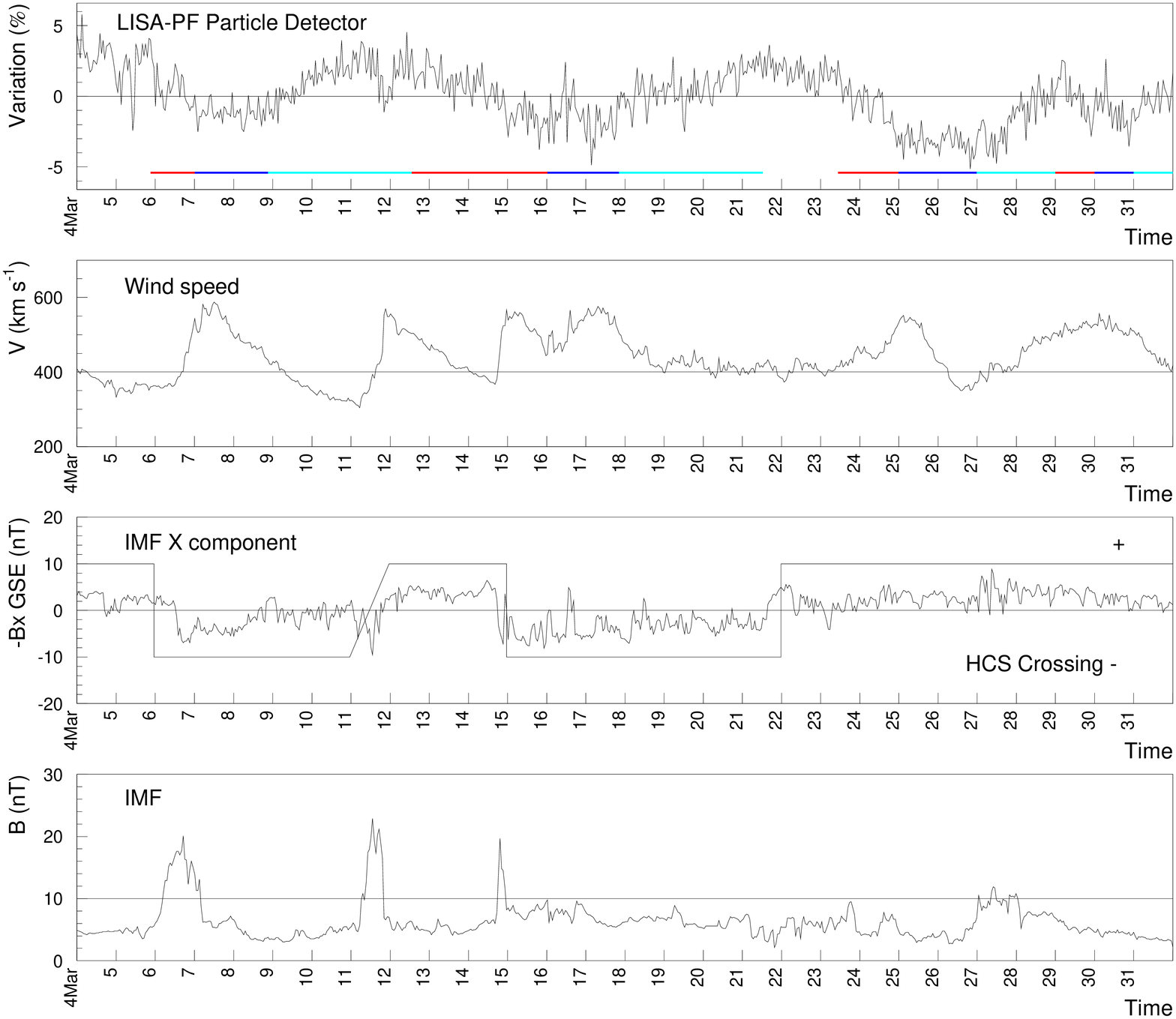}
  \caption{LPF PD counting rate fractional variations during the BR 2491 (March 4, 2016 - March 30, 2016)(first panel).  Solar wind speed (second panel), IMF radial component (third panel) and IMF intensity (fourth panel) contemporaneous measurements, gathered by the ACE experiment, are also  shown. HCS crossing (\url{http://omniweb.sci.gsfc.nasa.gov./html/polarity/polarity\_tab.html}) is shown in the third panel. Periods of time during which  the solar wind speed (V) and  the magnetic field (B) intensity remain below and above 400 km s$^{-1}$ and  10 nT, respectively, are shown in the second and fourth panels. Decrease, plateau and recovery periods of each GCR depression are represented by red, blue and cyan lines in the first panel.
}
  \label{figure1}
 \end{center}
\end{figure}
In the top panel of Fig. 6  the cosmic-ray flux fractional variations  during the BR 2491 (from March 4, 2016 through March 30, 2016) present four depressions (according to the definition reported above) starting on March 5, March 12, March 23 and March 29, respectively. The small deeps on March 11-12 and  March 19-20 along with  the small increase on March 16-17 are neglected. The solar wind plasma speed appears in the second panel of Fig. 6. The IMF radial component is shown  in the third panel and the IMF intensity in the fourth panel. In the third panel of Fig. 6 the HCS crossing (taken from \url{http://omniweb.sci.gsfc.nasa.gov./html/polarity/polarity\_tab.html}) is also indicated. Solar wind plasma and IMF data are taken from the ACE experiment. The  GCR flux depressions  appear associated with those periods of time during which  the solar wind plasma speed (V) is $>$ 400 km s$^{-1}$ and/or the IMF intensity (B) is $>$ 10 nT (second and fourth panels in Fig. 6).
This scenario basically corresponds to the passage of high-speed solar wind streams and/or CIRs \citep{harang,storini,cane,simpson,mckibben,bazi}.
When GCR short-term variations are correlated  with   the BV parameter,  the role of the magnetic field trend is privileged with respect to that of the solar wind speed since the IMF variations are larger than those of the solar wind speed. The   GCR depressions observed with LPF are associated with solar wind speed  changes smaller than 30\%  while the magnetic field is observed to increase up to a factor of 5.
Therefore,
a separate analysis of  B and V increases helps in better understanding the  dynamics of individual depressions
resulting from the interplay of several interplanetary structures which affect the role of different periodicities during the mission as observed in Figs 4 and 5.
From the point of view of time profiles  of individual depressions,  those   presenting similar durations for decrease and recovery phases are called $\it symmetric$. All the other depressions  are called $\it asymmetric$ \citep{budu}. The symmetric variations  are V or U shaped. 
Thirty-nine out of forty-five depressions were found asymmetric. Only six appeared symmetric, and out of these, five were found U-shaped and only one V-shaped.  
The period during which the  PD  counting rate remained at minimum values between decrease and recovery phases is called here $\it plateau$. 
A plateau is observed during both U-shaped symmetric and asymmetric depressions and appears correlated with the period the solar wind velocity remains above 400 km s$^{-1}$.  A typical asymmetric depression is that appearing in the top panel of Fig. 6 starting on March 5, 2016 with 2-day decrease, $\sim$ 1.5-day plateau and 3.5-day recovery periods.  In  the same figure a symmetric, U-cup-shaped depression starts on the 23rd of March with decrease, plateau and recovery phases lasting about 2 days each. Decrease, plateau and recovery phases for each depression during the Bartels rotation 2491 are shown in colors in the top panel of Fig. 6, as an example. Occurrence, characteristics and association with interplanetary  structures of all depressions are summarized in Table 2. The GCR flux depressions that commence at the interaction regions of slow and fast solar wind are associated to CIR  in Table 2. Depressions observed to commence during  different phases of corotating high speed solar wind stream passage  are indicated by CHSS. HCS crossing (HCSC) and ICMEs are observed to play a minor role in modulating the GCR flux with respect to corotating high speed solar wind streams during the LPF mission.  In Table 2 MFE (magnetic field enhancement) indicates a magnetic structure present in the slow solar wind. The association among magnetic structures and GCR flux depressions was carried out on the basis of contemporaneous interplanetary magnetic field and solar wind parameter observations from the ACE experiment. 
 
\startlongtable
\begin{deluxetable}{c|c|c|c|c|c|c}
\tablecaption{Occurrence and characteristics of the GCR flux depressions observed during the LPF mission\label{tab:table}. Interplanetary structures associated with the depressions are indicated (CIR: corotating interaction region; CHSS: corotating high-speed solar wind streams; ICME: interplanetary coronal mass ejection (\url{http://www.srl.caltech.edu/ACE/ASC/DATA/level3/icmetable2.htm}); S: shock; MC: magnetic cloud; HCSC: heliospheric current sheet crossing; MFE: magnetic field enhancement in the slow solar wind).}
\tablehead{
  \colhead{Date} & \colhead{Onset} & \colhead{Decrease}&\colhead{Plateau}&\colhead{Recovery}& \colhead{Amplitude}& Interplanetary structure\\
\colhead{} & \colhead{Time} & \colhead{Days} & \colhead{Days} & \colhead{Days} & \% &
}
\startdata
February 26, 2016 & 16.00 UT & 2.5 & 1.0 & 3.2 & 7.0 & CHSS\\
March 5, 2016 & 21.00 UT     & 2.0 & 1.0 & 3.5 & 4.9 & ICME+CHSS\\
March 12, 2016 & 00.00 UT    & 3.5 & 2.0 & 3.5 & 5.3 & CHSS\\
March 23, 2016 & 11.00 UT    & 2.0 & 2.0 & 2.0 & 6.0 & CHSS\\
March 29, 2016 & 03.00 UT    & 1.0 & 1.0 & 4.0 & 3.4 & CHSS\\
April 10, 2016 & 11.00 UT    & 4.0 & 0.0 & 5.5 & 3.1 & MFE+HCSC+ICME\\
April 20, 2016 & 12.00 UT    & 3.0 & 2.0 & 4.5 & 7.1 & CHSS\\
May 1, 2016 & 11.00 UT    & 1.5 & 1.0 & 2.0 & 2.8 & CHSS\\
May 6, 2016 & 00.00 UT   & 2.8 & 0.0 & 6.5 & 4.7 & CHSS\\
May 15, 2016 & 12.00 UT  & 4.0 & 1.0 & 1.0 & 7.2 & CIR\\
May 29, 2016 & 13.00UT      & 1.5 & 0.0 & 5.0 & 3.0 & CHSS\\
June 5, 2016 & 04.00 UT      & 1.0 & 1.0 & 4.0 & 5.1 & CIR\\
June 12, 2016 & 07.00 UT      & 3.5 & 0.0 & 10.0 & 8.4 & CHSS\\
June 30, 2016 & 07.00 UT     & 1.0 & 4.0 & 4.5 & 2.5 & MFE+HCSC\\
July 7, 2016 & 00.00 UT      & 6.0 & 1.0 & 3.0 & 11.9 & CIR\\
July 20, 2016 & 07.00 UT     & 1.0 & 1.0 & 12.0 & 5.4 & ICME+CHSS\\
August 2, 2016 & 12.00 UT    & 1.0 & 0.0 & 2.8 & 9.0 & ICME (S,MC)+ CHSS\\
August 5, 2016 & 21.00 UT    & 5.0 & 4.0 & 15.0 & 6.8 & CHSS\\
August 29, 2016 & 21.00 UT   & 6.0 & 2.0 & 19.0 & 8.6 & CIR\\
September 26, 2016& 12.00 UT & 3.0 & 2.0 & 8.0 & 6.9 & CIR\\
October 11, 2016 & 15.00 UT  & 2.0 & 0.0 & 2.0 & 5.0 & HCSC+ICME\\
October 16, 2016 & 15.00 UT  & 1.0 & 0.0 & 5.5 & 2.8 & CHSS\\
October 23, 2016 & 00.00 UT  & 6.0 & 2.0 & 8.0 & 7.5 & CHSS\\
November 12, 2016 & 00.00 UT & 1.0 & 3.0 & 4.0 & 1.6 & CIR\\
November 20, 2016 & 16.00 UT & 5.0 & 3.5 & 5.0 & 8.1 & HCSC+CHSS\\
December 5, 2016 & 00.00 UT  & 1.0 & 1.0 & 1.0 & 1.9 & MFE\\
December 7, 2016 & 12.00 UT  & 2.0 & 3.5 & 4.5 & 2.8 & CIR\\
December 17, 2016 & 19.00 UT & 8.5 & 1.0 & 4.5 & 10.9 & CHSS\\
January 5, 2017 & 03.00 UT   & 1.0 & 2.0 & 6.5 & 3.0 & CIR\\
January 14, 2017 & 15.00 UT  & 5.0 & 2.0 & 2.0 & 6.3 & HCSC + CHSS\\
January 25, 2017 & 11.00 UT  & 2.5 & 0.0 & 3.0 & 3.4 & HCSC + CHSS\\
January 30, 2017 & 16.00 UT  & 3.0 & 0.0 & 10.5 & 4.4 & CIR\\
February 16, 2017 & 23.00 UT & 1.0 & 1.0 & 2.5 & 3.1 & CIR\\
February 23, 2017 & 10.00 UT  & 1.0 & 0.0 & 3.0 & 1.8 & CIR\\
March 1, 2017 & 05.00 UT     & 2.0 & 0.0 & 5.5 & 3.9 & CIR\\
March 21, 2017 & 00.00 UT    & 2.0 & 1.0 & 2.0 & 4.4 & CIR\\
March 27, 2017 & 00.00 UT    & 8.0 & 3.5 & 5.5 & 6.9 & CIR\\
April 18, 2017 & 09.00 UT    & 1.5 & 0.5 & 1.0 & 3.4 & CIR\\
April 21, 2017 & 11.00 UT    & 3.0 & 2.0 & 4.5 & 7.8 & CIR\\
May 1, 2017 & 00.00 UT       & 1.5 & 1.0 & 11.0 & 1.3 & HCSC\\
May 15, 2017 & 08.00 UT      & 1.5 & 1.0 & 2.0 & 3.8 & CIR\\
May 19, 2017 & 10.00 UT      & 1.5 & 1.0 & 5.5 & 2.5 & HCSC+CHSS\\
May 27, 2017 & 18.00 UT      & 1.0 & 1.0 & 9.0 & 5.6 & ICME (S)\\
June 12, 2017 & 16.00 UT      & 6.0 & 2.0 & 2.5 & 3.4 & CHSS\\
June 24, 2017 & 14.00 UT     & 4.0 & 1.0 & 2.0 & 5.3 & CIR\\
\enddata
\end{deluxetable}

Average durations of decrease, plateau and recovery periods for the forty-five GCR depressions observed with LPF are reported in Table 3.

\begin{table}
\centering
\caption{\label{table3} Average characteristics of GCR flux depressions observed with LPF.}
\begin{tabular}{lll}
\hline                      
\hline
 &Duration & \\
 & (Days) & (\%)  \\
\hline
Decrease &  2.8 $\pm$ 2.0   &\\
Plateau  &  1.3 $\pm$ 1.2&\\
Recovery &  5.1 $\pm$ 3.8 &\\
Total duration   &  9.2 $\pm$ 5.0 & \\
Intensity  &   & 5.1 $\pm$ 2.5 \\
\hline
\end{tabular}
\end{table}

The average GCR flux depression amplitude  of 5.1$\pm$2.5\%  appears consistent, within statistical errors, with that reported by  \citet{richar} of 3.2$\pm$0.1\%
 for  particle nominal  energies larger than 60 MeV.  The   cut-off energy of particles observed with Helios I, Helios 2 and IMP8  was poorly estimated \citep{richar} while for the  LPF PD observations the same was set with both Monte Carlo simulation and beam test experiment \citep{ara,mateos}. However, since the majority of cosmic-ray particles lie in the energy range of hundreds of MeV, a slightly difference in the detection capability of  low energy particles  is not expected to make a relevant difference for the above comparison.

The full evolution of one classical, two-step Forbush decrease \citep{cane1}
 was detected aboard LPF on August 2, 2016 as it is shown in  Fig. 7.  In this figure all panels are the same as in Fig. 6.
\begin{figure}
  \begin{center}
  \includegraphics[width=0.7\textwidth]{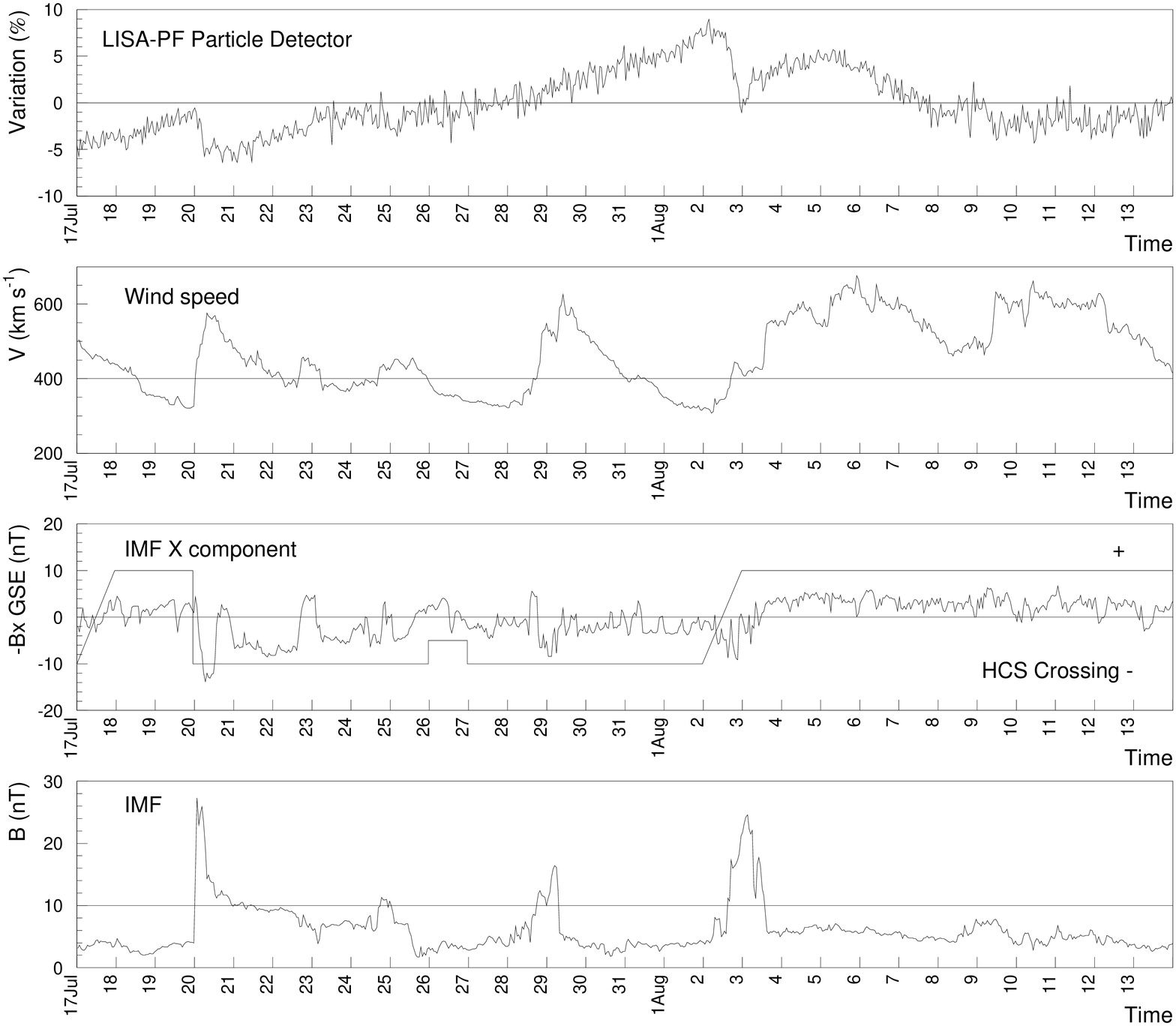}
  \caption{Same as Fig. 6 for the BR 2496 (July 17, 2016 - August 12, 2016).}
  \label{figure1}
 \end{center}
\end{figure}
The sharp decrease of the GCR flux on the 2nd of August  lasted about 10 hours after 12.00 UT, no plateau was observed and the recovery period was modulated by an incoming high speed solar 
wind stream. The GCR depression appears correlated with a contemporaneous increase of the interplanetary magnetic field intensity up to 24 nT, while the solar wind speed barely passed  400 km s$^{-1}$.



\section{Energy-dependence of GCR flux short-term  depressions} \label{sec:highlight}

Some clues are reported in the literature about the energy dependence of  GCR flux short-term depressions.
The energy-dependence of 27-day  GCR flux variations, for instance, is discussed in \citet{gri15} and references therein. 
 The shielding effect of the atmosphere and the geomagnetic cut-off prevent neutron monitors to carry out a direct measurement of cosmic-ray energy spectra below effective energies of several GeV, although they can be obtained by using models combined with neutron monitors observations \citep{fluck,beer,uso11,uso17}.
Interesting  attempts to investigate the energy-dependence of short-term depressions of cosmic-ray fluxes above a few tens of MeV, through direct measurements with  magnetic spectrometers,   were carried out by the balloon-borne experiment BESS-Polar I \citep{bessp} and the satellite  experiment PAMELA  \citep{pamforb}.
BESS-Polar I flew from Williams Field near Mc Murdo Station from December 13, 2004  through December 21, 2004. At the beginning of the 
flight, this balloon-borne experiment observed a recovering proton flux from a previous decrease. The recovery intensity appeared to be of
 8-9\% below 0.86 GeV and of 3\% above 6 GeV. The authors claimed that this occurrence was due to the transit of a CIR interface  or a 
magnetic cloud  \citep{burlaga} or a combination of the two. This experiment detected a new GCR proton flux depression after the passage of a high-speed solar wind stream on December 17, 2004.
The PAMELA experiment carried out the first measurement of proton and helium nucleus differential fluxes  in space  during  a  Forbush decrease on  December 14, 2006 (16.50 UT-22.35 UT) after two SEP events dated December 13, 2006 and December 14, 2006. Unfortunately, balloon-borne
 magnetic spectrometer  experiments, like BESS-Polar I,  have short duration and space-borne instruments, like PAMELA,  have small geometrical factors and
therefore data must be integrated over periods longer than the typical one-hour data binning required to study recurrent GCR depressions.
\begin{deluxetable}{llll}
  \tablecaption{Neutron monitor station location and characteristics.\label{tabneut}}
  \tablehead{
    \colhead{} & \colhead{Location} & \colhead{Vertical} & \colhead{} \\
    \colhead{} &  \colhead{} &\colhead{cut-off} & \colhead{Effective}\\
    \colhead{Station} &  \colhead{} & \colhead{rigidity} & \colhead{Energy}\\
    \colhead{} &  \colhead{} & \colhead{GV} & \colhead{GeV}
  }
  \startdata
  Thule & North Pole  & 0.3  &  11-12  \\
  Terre Adelie & South Pole & 0.0   & 11-12     \\
  Mc Murdo & South Pole & 0.3   & 11-12     \\
  Oulu &  Finland & 0.8   &  12    \\
  Rome    & Italy  & 6.3  &  17   \\
  Mexico & Mexico & 8.2  &  20   \\
  \enddata
\end{deluxetable}

The GCR flux fractional variations observed with the LPF PD  have been compared to contemporaneous similar measurements carried out with  neutron monitors placed at different geographic latitudes. Location, vertical cut-off rigidities and effective energies for all neutron monitor stations considered in this work are reported in Table 4. Both LPF PD and neutron monitor  data were hourly averaged and appear  in Figs. 8 and 9 for the BRs 2491 and 2496, respectively, as an example.  This comparison indicates that while the maximum and average  GCR fractional variations observed with LPF above 70 MeV n$^{-1}$ are of more than 11\% and of about  5\%, respectively, the same goes down to a maximum of 3\% above 11-12 GeV in near-polar stations and to a maximum of 2\% above 15 GeV at increasing latitudes.
\begin{figure}
  \begin{center}
  \includegraphics[width=0.7\textwidth]{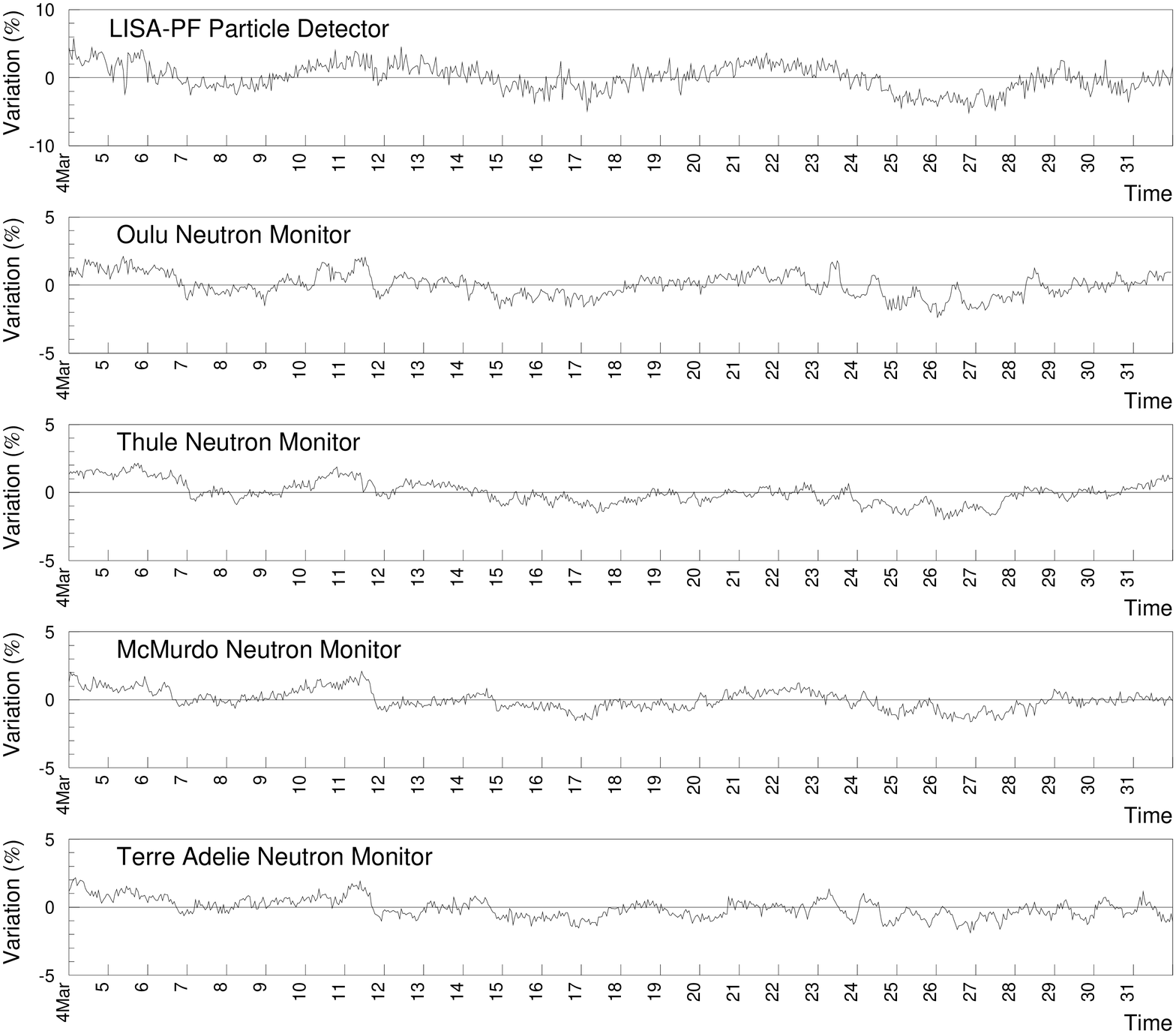}
  \caption{Comparison of LPF PD counting rate fractional variations with contemporaneous, analogous measurements of polar neutron monitors  during the BR 2491 
(March 4, 2016 - March 31, 2016).
}
  \label{figure1}
 \end{center}
\end{figure}

\begin{figure}
  \begin{center}
  \includegraphics[width=0.7\textwidth]{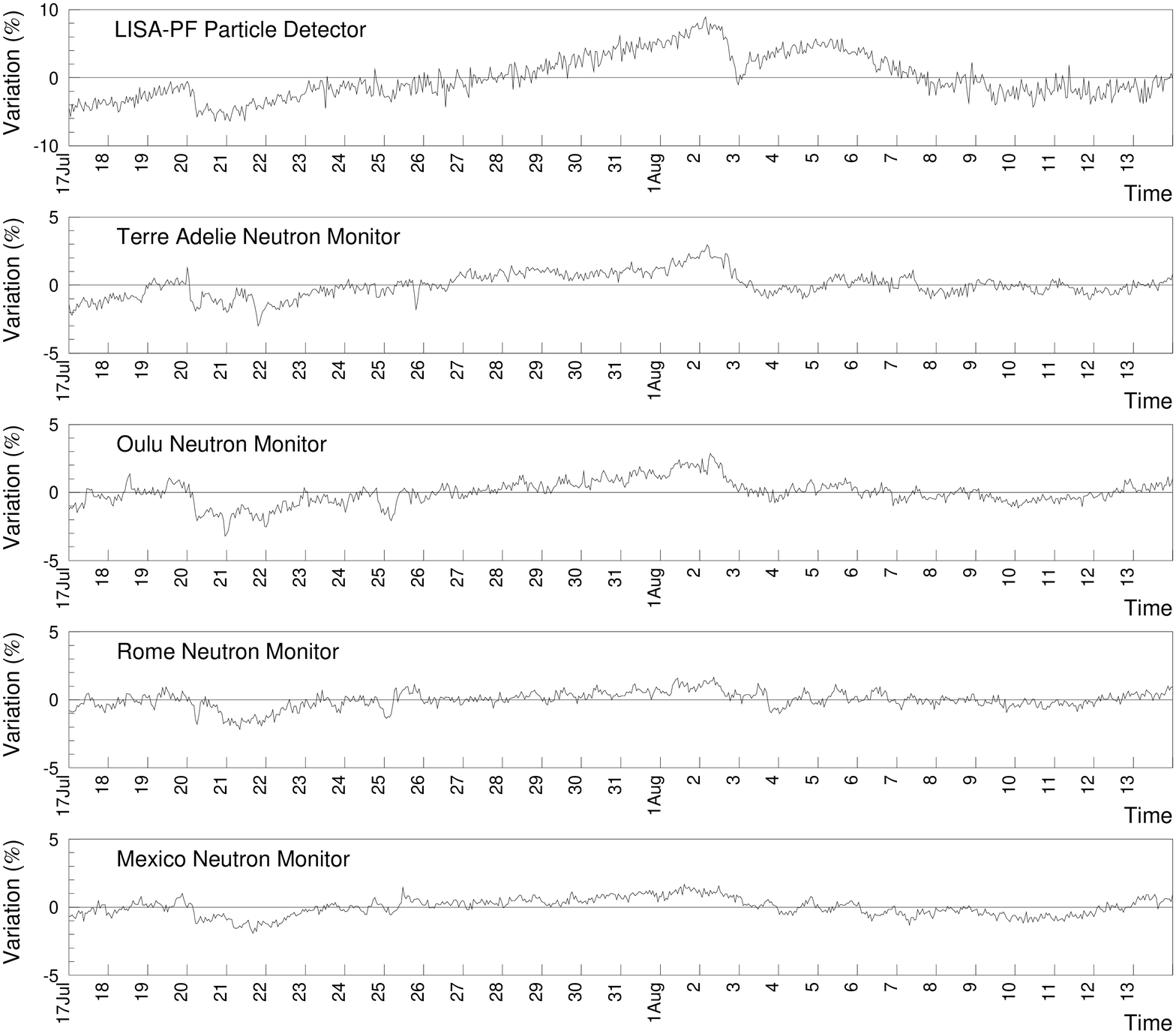}
  \caption{ Comparison of LPF PD counting rate fractional variations with contemporaneous, analogous measurements of neutron monitors placed at various geographic latitudes during the BR 2496 (July 17, 2016 - August 13, 2016). 
}
  \label{figure1}
 \end{center}
\end{figure}  
 
 During the Forbush decrease  observed on August 2, 2016 on LPF,  a 3-$\sigma$ decrease of the GCR flux occurred between 12.00 UT and 16.00 UT. The GCR flux reached its minimum at 22.40 UT: data indicated a GCR flux fractional decrease of  9\% in L1 in  $\sim$ 10 hours. The GCR flux depression  recovered soon after the deep.
The trend of this GCR flux depression appears different from recurrent GCR flux variations  observed to be  of 2\% - 3\% day$^{-1}$ and  1\% - 2\% day$^{-1}$ during the decrease and recovery periods, respectively. 
In Fig. 9 it can be noticed that the amplitude of the same depression is found to be
of 3\% in near-polar Terre Adelie and Oulu stations while it is  of just 2\% and 1\% in  Rome and Mexico stations, respectively. 
In order to determine the GCR proton energy differential flux  at the deep of the depression at 22.40 UT on LPF, the proton energy differential flux for the month of August 2016 ($\phi$=438 MV) was estimated  first above 70 MeV and parameterized following equation 3, as described in Section 3. The proton integral flux in August 2016  was then calculated as a integral of this differential flux. 
The proton integral flux, thus obtained, was  properly reduced at 70 MeV as indicated by the LPF PD data and at the  effective energy of each neutron monitor (reported in Table 4) on the basis of the neutron monitor decreases. Finally, the differential flux was inferred from the integral flux.
The proton differential flux in August 2016 before the Forbush decrease  and that estimated at the deep of the depression at 22.40 UT on August 2, 2016 were  parameterized as reported in  Table 5 and are compared in Fig. 10.

  The  helium  differential flux  at the deep of the depression was not estimated  since no accurate proton-helium separation was allowed by the PD aboard LPF and the data trend is biased by protons since the He/p ratio in GCRs is about 0.1.

Measurements of the energy dependence of GCR flux recurrent and non recurrent depressions  and the study of their evolution can be used to estimate the test-mass charging aboard future generation LISA-like interferometers  \citep[][and references therein]{gri15}. 
 Despite minor  changes in the  instrument performance due to GCR short-term variations are expected, future  interferometers devoted to gravitational wave detection in space will detect sub-femto-g spurious acceleration at low frequencies ($\sim$ $10^{-5}$ Hz), and  the role of any  interplanetary disturbance  must be evaluated and quantified. 



\begin{table}
\centering
\caption{\label{table4} Parameterizations of proton energy spectra for August 2016 before (continuous line in Fig. 10) and at the deep of the GCR depression observed on the 2nd of August (dotted line in Fig. 10).}
\begin{tabular}{lllll}
\hline
\hline
&  $A$  &  $b$ &   $\alpha$  &  $\beta$  \\
\hline
p (August\ 2016)  & 18000 & 1.01 & 3.66 & 0.87\\
p (August\ 2nd,\ 2016; depressed)  & 18000 & 1.068 & 3.66 & 0.869\\
\hline
\end{tabular}
\end{table}



\begin{figure}
  \begin{center}
  \includegraphics[width=3.5in]{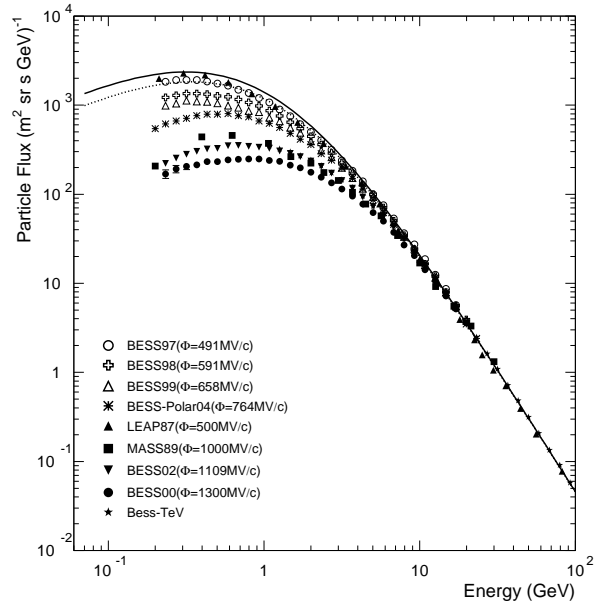}
  \caption{GCR proton  energy spectra measurements and projections before (continuous line) and at the deep (dotted line) of the depression observed on August 2, 
2016 with LPF.
}
  \label{figure1}
 \end{center}
\end{figure}

\section{Capability of the LPF PD in monitoring the passage of interplanetary coronal mass ejections} \label{sec:cite}

In this Section  it is evaluated the capability of space missions like LPF, carrying PDs  optimized for GCR detection,  in  monitoring the passage of ICMEs and in forecasting geomagnetic activity, when these interplanetary structures present intense southward magnetic field that reconnect with the Earth magnetic field and induce geomagnetic activity.

In Fig. 11   the  Forbush decrease observed with LPF on the 2nd of August 2016 is compared to the contemporaneous IMF intensity and solar wind speed measured by ACE.
The transit of an ICME near Earth from August 2, 2016 at 14.00 UT and August 3, 2016 at 3.00 UT, is indicated in the same figure by dashed lines (\url{http://www.srl.caltech.edu/ACE/ASC/DATA/level3/icmetable2.htm}; \citet[see also][]{ricane}). A detailed description of the characteristics of this ICME is reported in 
\url{http://www.stce.be/esww14/contributions/public/S4-P1/S4-P1-08-BenellaSimone/Poster_ESWW.pdf}.
\begin{figure}
  \begin{center}
  \includegraphics[width=3.5in,,angle=90]{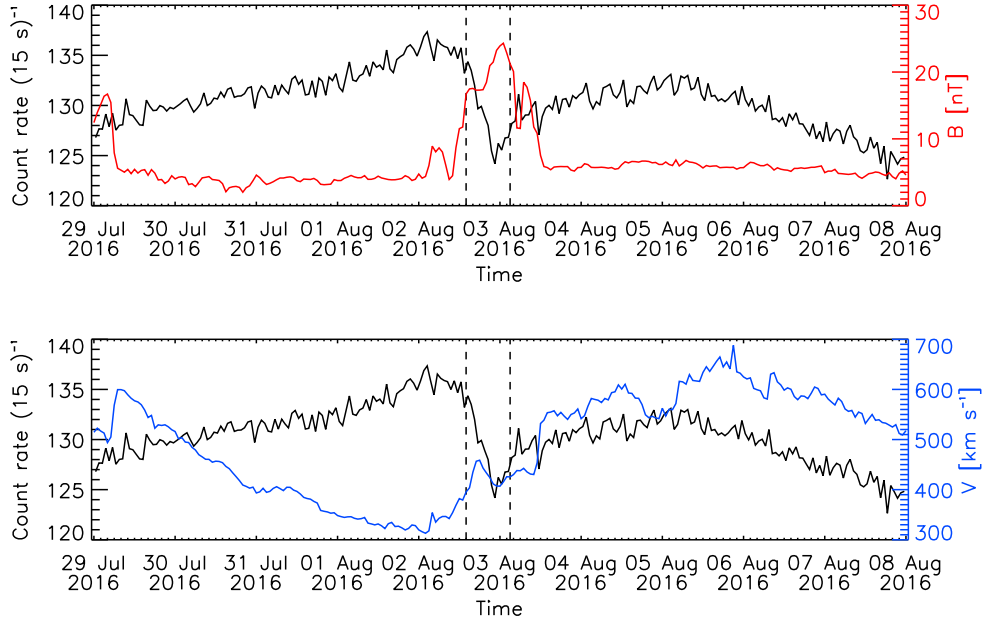}
  \caption{Comparison of the LPF PD counting rate fractional variations with IMF intensity (red line, right scale in the top panel) and solar wind speed (blue line right scale in the bottom panel) between July 29, 2016 and August 8, 2016. 
The vertical dashed line represent the beginning of the GCR flux depression observed with LPF and the passage of an ICME.
}
  \label{figure1}
 \end{center}
\end{figure}

\begin{figure}
  \begin{center}
  \includegraphics[width=5.in]{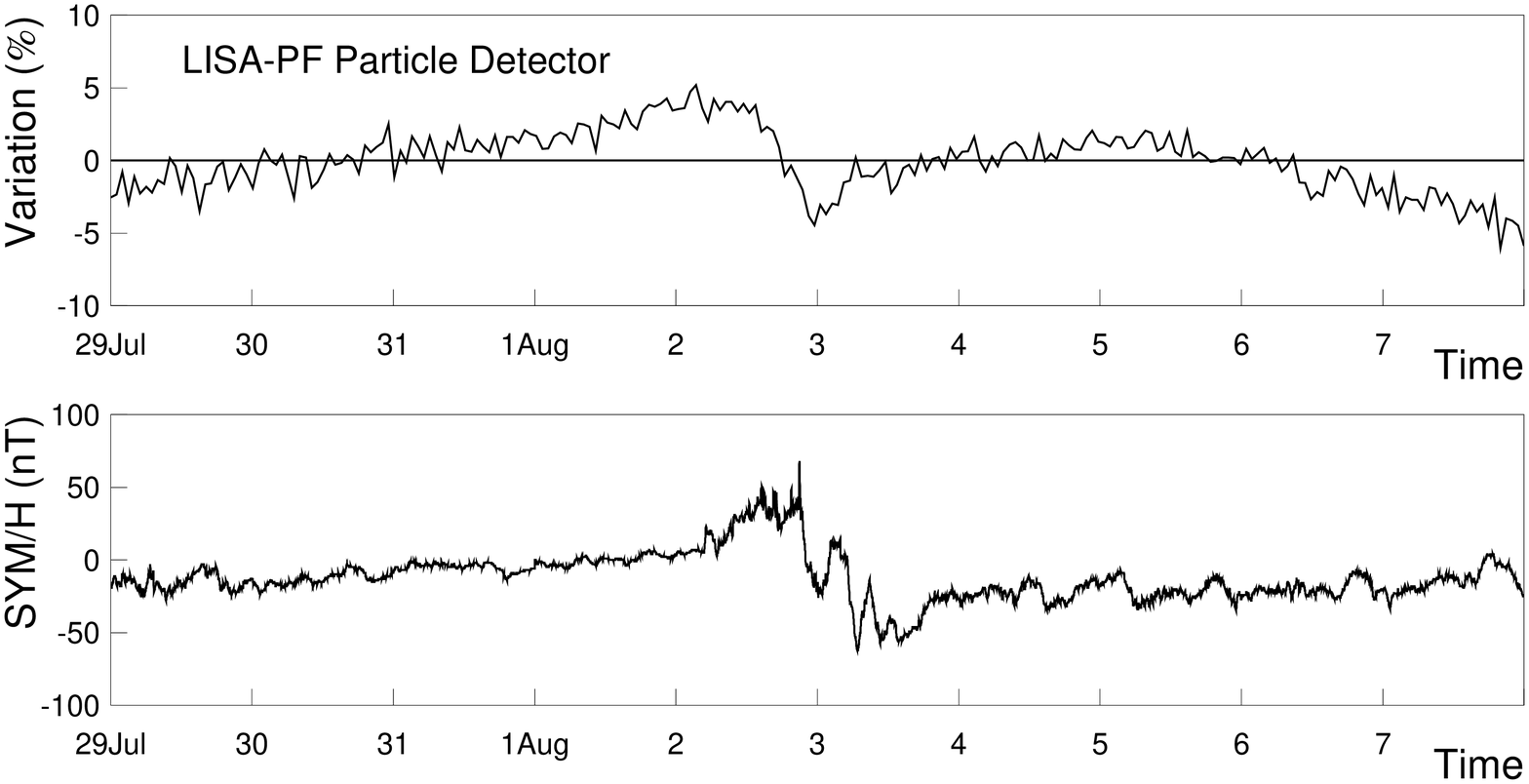}
  \caption{Comparison of the LPF PD counting rate fractional variations with the geomagnetic index SYM-H
for the same period indicated in Fig. 11.}
  \label{figure1}
 \end{center}
\end{figure}

The decrease phase of the cosmic-ray flux seems to occur in two steps, suggesting that LPF crossed the region of the shock of the ICME and then the ejecta \cite[][]{cane1}. 
A geomagnetic disturbance was observed to start (K$_p >$ 5) after 21.30 UT. In Fig. 12 the LPF GCR flux fractional variations are also compared to the SYM-H geomagnetic index which allows to  follow the evolution of a geomagnetic disturbance at low latitudes. 
The characteristics of each GCR short-term flux depression are unique, often resulting from the interplaying effects of consecutive structures propagating in the interplanetary medium. 
However, for the August 2, 2016  Forbush decrease, an alert issued by LPF at the time  a 3-$\sigma$ GCR flux decrease was reached around 16.00 UT,  would have anticipated the  geomagnetic disturbance observed at the Earth by  several hours in case of appropriate baseline communication strategy. PDs aboard space missions allow for studying the energy dependence of GCR short-term depressions, their evolution and association with interplanetary  structures better than allowed by the use of neutron monitor measurements only \citep[see also][and references therein]{cane1}. The ICME tracking in space by Forbush decreases was also recently discussed in \citet{witasse}. 
\section{Conclusions}

A PD aboard the ESA mission LPF allowed for the study of GCR short-term flux depressions  
above 70 MeV n$^{-1}$ during the descending phase of the solar cycle N$^{\circ}$ 24. The majority of these depressions are recurrent and associated with corotating high speed solar wind streams.  ICMEs and  heliospheric current sheet crossing play a minor role. The average duration of GCR flux depressions  observed aboard LPF are found of 
9.2$\pm$5.0  days. Decrease, plateau and recovery average periods are 2.8$\pm$2.0 days, 1.3$\pm$1.2 days and 5.1$\pm$3.8  days, respectively.
The average depression intensity is 5.1$\pm$2.5\%. 

The proton energy differential flux at the deep of  a Forbush decrease observed on August 2, 2016 was obtained from the integral  energy spectrum measurements 
carried out   with  LPF PD data and from those of neutron monitors placed in sites characterized by an increasing effective energy. Finally, it was shown that 
LISA-like and other missions in space, even if primarily devoted to difference science investigations, in some cases,  may play the role of sentinels in monitoring the passage of magnetic structures that, when characterized by intense southern components of the magnetic field, induce geomagnetic activity.


\acknowledgments
The authors are grateful to the anonymous reviewer for his/her constructive comments and precious suggestions which helped us to greatly improve the manuscript.
D. Telloni  is financially supported by the Italian Space Agency (ASI) under contract I/013/12/0. Sunspot number  and solar modulation parameter data were gathered from 
\url{http://www.sidc.be/silso/home} and \url{http://cosmicrays.oulu.fi/phi/Phi\_mon.txt}, respectively. Data from Wind and ACE experiments were obtained from the NASA-CDAWeb website. We acknowledge the NMDB database (www.nmdb.eu) funded under the European Union's FP7 programme (contract no. 213007), and the PIs of individual neutron monitors for providing data.  HCS crossing was taken from \url{http://omniweb.sci.gsfc.nasa.gov./html/polarity/polarity\_tab.html}.
This work has been made possible by the LISA Pathfinder mission, which is part of the space-science program of the European Space Agency. The French contribution has been supported by the CNES (Accord Specific de projet CNES 1316634/CNRS 103747), the CNRS, the Observatoire de Paris, and the University Paris-Diderot. E. P. and H. I. also acknowledge the financial support of the UnivEarthS Labex program at Sorbonne Paris Cit\'e (ANR-10-LABX-0023 and ANR-11-IDEX-0005-02). The Albert-Einstein-Institut acknowledges the support of the German Space Agency, DLR. The work is supported by the Federal Ministry for Economic Affairs and Energy based on a resolution of the German Bundestag (FKZ 50OQ0 501 and FKZ 50OQ1601). The Italian contribution has been supported by Agenzia Spaziale Italiana and Instituto Nazionale di Fisica Nucleare.
 The Spanish contribution has been supported by Contracts No. AYA2010-15709 (MICINN), No. ESP2013-47637-P, and No. ESP2015-67234-P (MINECO). M. N. acknowledges support from Fundacion General CSIC (Programa`ComFuturo). F.R. acknowledges an FPI contract (MINECO). The Swiss contribution acknowledges the support of the Swiss Space Office (SSO) via the PRODEX Program of ESA. L. F. acknowledges the support of the Swiss National Science Foundation. The United Kingdom groups acknowledge support from the United Kingdom Space Agency (UKSA), the University of Glasgow, the University of Birmingham, Imperial College, and the Scottish Universities Physics Alliance (SUPA). J. I. T. and J. S. acknowledge the support of the U.S. National Aeronautics and Space Administration (NASA). N. Korsakova acknowledges the support of the Newton International Fellowship from the Royal Society.

\end{document}